\documentclass[aip,amsmath,amssymb,reprint]{revtex4-2}
\usepackage[normalem]{ulem}
\usepackage[utf8]{inputenc}
\usepackage[english]{babel}
\usepackage{amsmath}
\usepackage{amsfonts}
\usepackage{amssymb}
\usepackage{figsize}
\usepackage{graphicx}
\usepackage[dvipsnames]{xcolor}
\usepackage{bm}
\usepackage{xcolor}
\usepackage{soul}
\usepackage{comment}
\usepackage{appendix}
\usepackage[colorlinks=true, allcolors=blue]{hyperref}

\newcommand{\jd}[1]{\textcolor{purple}{JD: #1}}

\begin{document}
\title{Free energies and optimal reaction coordinates via entropy production}

\author{Jérémy Diharce}
\affiliation{Sorbonne Université, Musée National d’Histoire Naturelle, UMR CNRS 7590, Institut de
Minéralogie, de Physique des Materiaux et de Cosmochimie, IMPMC, F-75005 Paris, France}
\author{Line Mouaffac}
\affiliation{Sorbonne Université, UMR 8234, Laboratoire de Physico-chimie des Electrolytes et Nanosystèmes Interfaciaux, PHENIX, F-75005 Paris, France}
\author{Axel Dian}
\author{Fabio Pietrucci}
\email{fabio.pietrucci@sorbonne-universite.fr}
\affiliation{Sorbonne Université, Musée National d’Histoire Naturelle, UMR CNRS 7590, Institut de
Minéralogie, de Physique des Materiaux et de Cosmochimie, IMPMC, F-75005 Paris, France}

\begin{abstract}
We show through theory and numerical experiments that a straightforward calculation of entropy production on short molecular dynamics trajectories allows estimating free-energy barriers and identifying optimal reaction coordinates of activated processes.
To this aim, we perform an analysis based on stochastic energetics on a set of trajectories relaxing towards equilibrium from a same initial configuration, projected on different putative coordinates.
After demonstrating the approach on simple benchmarks, 
we show that it is possible to estimate the free-energy barrier of a complex high-dimensional system (carbon nanoparticles in water) and to rank the quality of order parameters, in agreement with the committor probability.
The results shed light on the entanglement between the second law, free-energy landscapes, reaction coordinates and kinetic rates.
\end{abstract}

\maketitle


Recent years have seen intense research and important advances in two scientific fields, unrelated at first sight: the study of entropy production in irreversible thermodynamics, and the quest for optimal order parameters in rare events.
The first field is mainly concerned with non-equilibrium conditions, like in externally-driven systems and active matter, the second with unperturbed equilibrium systems where thermal fluctuations induce transition processes, like phase transitions, chemical reactions or protein folding.

In this work we show how theoretical tools for the estimation of entropy production from trajectories, recently developed by Sekimoto~\cite{Sekimoto09} and Seifert~\cite{Seifert05}, can be fruitfully borrowed to solve the very hard problem of identifying optimal order parameters for the transformations of matter, equivalent to the committor function $\varphi$ (i.e., the probability of an atomic configuration to evolve reaching the product state before the reactant state), without resorting to the brute-force estimation of the latter, of prohibitive cost~\cite{Mouaffac23,Jung23}.

Langevin equations provide a suitable mathematical framework to approximate in a statistically-faithful way the time evolution of a complex, high-dimensional system, projected on a generalized coordinate $x$. 
The latter is explicitly included in the equation of motion, while all other degrees of freedom are implicitly collected into a thermal bath, whose effects on the explicit coordinate are summarized by friction and by a random force.
Sekimoto elegantly endowed Langevin trajectories with precise definitions of heat and work, thus connecting with the first law of thermodynamics, while Seifert discovered a remarkably simple formula to compute the rate of entropy production along such trajectories, thus connecting with the second law.

Two important aspects need to be stressed. First, the relaxation towards equilibrium of a system in contact with a thermal bath can be characterized as an increase of the total entropy $S_{tot}=S+S_{bath}$, sum of the entropies of the system and of the bath, until reaching the maximum value.
Second, the total entropy production is formally equal to a drop in free energy, since the energy variation of the system is, in fact,  heat reversibly exchanged with the bath:
\begin{equation}\label{eq:Stot_def}
\Delta F=\langle \Delta E\rangle -T\Delta S \equiv -T\Delta S_{bath}-T\Delta S\equiv-T\Delta S_{tot}
\end{equation}
where $\beta\Delta F=-\log(P_{end}/P_{begin})$ is given by the ratio between the probabilities of the final and initial states, not to be confused with a mere "altitude" difference in a free-energy landscape. 

To connect the previous theory with the committor, generally considered as the optimal reaction coordinate for a transformation linking two metastable states $A$ and $B$, we consider the following experiment: a system is prepared in a local-equilibrium state constrained in the vicinity of a transition state ($\varphi=1/2$); the constraint is then removed, so that the system relaxes spontaneously to local equilibrium in either A or B.
If we focus on B, the relaxation produces 
$\Delta S_{tot}/k_B=-\log (P_{TS}/P_B)$. 
Customarily, probabilities are estimated via order parameters: observing the relaxation using the committor -- by definition able to resolve the $TS$ from $B$ -- as order parameter,  grants a correct estimate of the probability ratio, hence of $\Delta F$ and of $\Delta S_{tot}$.
However, other variables overlap in the same region the $TS$ with other states (while $B$ is generally easier to resolve), thus overestimating $P_{TS}/P_B$ and underestimating $\Delta S_{tot}$ (see Fig. 1).
The committor is therefore the "most irreversible" order parameter, i.e., the one manifesting the largest entropy production.

\begin{figure}[h!]
    \centering
    \includegraphics[width=0.4\textwidth]{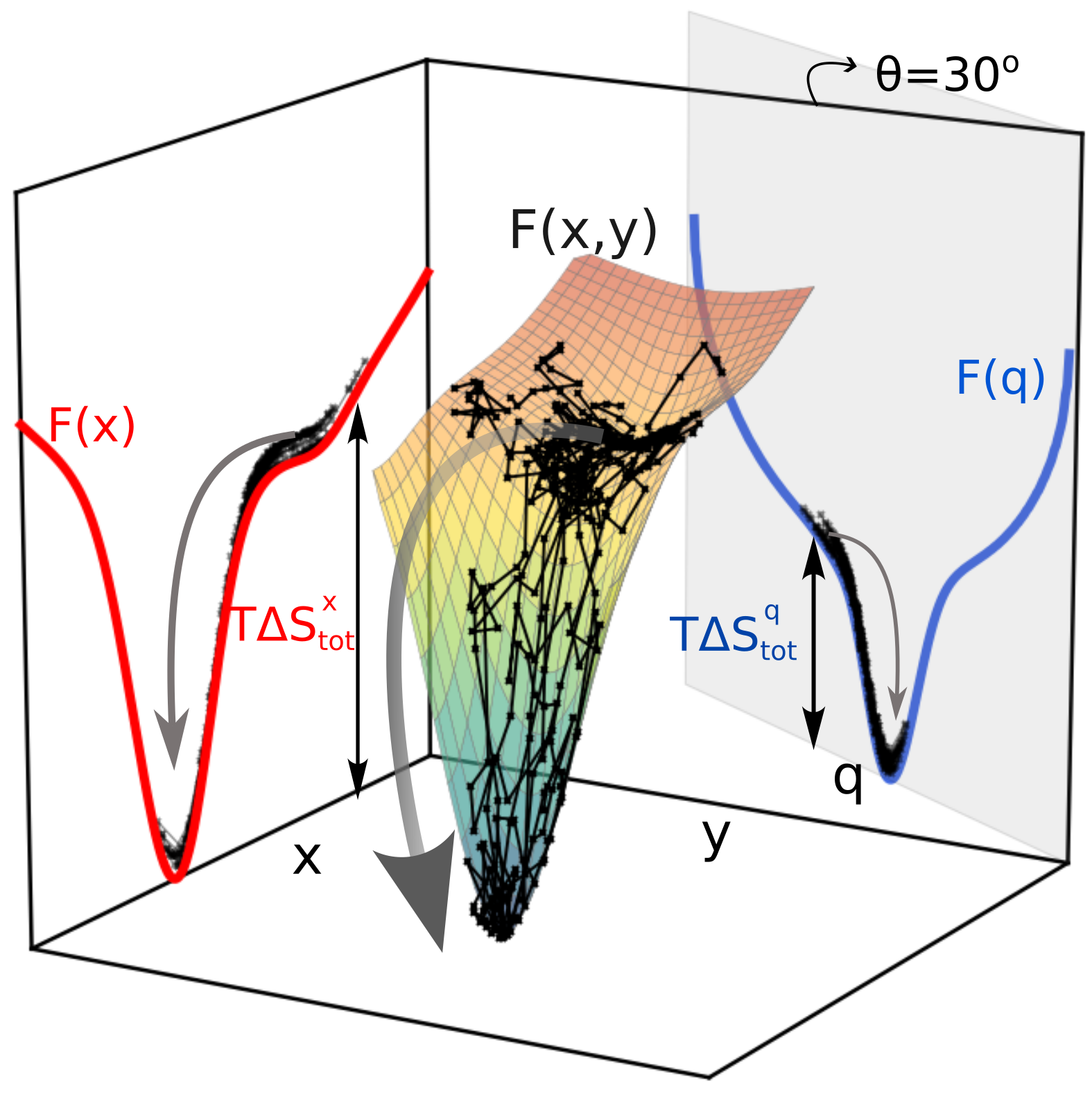}
    \caption{
    Illustrative example of the connection between relaxation processes, entropy production, and free-energy differences.
    Several trajectories are shown, evolving on a free-energy landscape $F(x,y)$ from an initial state (high free energy) to a final state (low free energy).
    According to thermodynamics, any spontaneous relaxation process entails an increase in total entropy $T\Delta S_{tot}=-\Delta F$.
    However, when observed from the viewpoint of a generic single variable $q$, the free-energy drop and the entropy production appear reduced from the true value, due to properties of the marginal probability density $\rho(q)=e^{-\beta F(q)}/Z$, unless an optimal projection coordinate (here $x$) is employed.
    }
    \label{fig:fig1}
\end{figure}

This observation opens the possibility of optimizing the order parameter, starting from a basis set of generalized coordinates, by identifying the definition that maximizes $\Delta S_{tot}$ over a given set of trajectories relaxing between the same initial and final states.
Clearly, the procedure would also yield the correct $\Delta F$ value.
The approach turns out to be numerically straightforward thanks to a formula introduced by 
Seifert, hereafter dubbed entropy production rate integral (EPRI), yielding $\Delta S_{tot}$ for overdamped Langevin trajectories along a variable $x$ (hereafter, we consider $k_B=1$ for simplicity):

\begin{equation}\label{eq:Seifert1}
\begin{split}
\Delta S_{tot} & =\int_0^{t_{max}}\dot{S}_{tot}\,dt
=\int_0^{t_{max}} dt\,\frac{\langle v^2\rangle}{D} \\
& =\int_0^{t_{max}} dt\int dx\,\frac{\rho(x,t)\,v(x,t)^2}{D(x)}
\end{split}
\end{equation}
where $\rho(x,t)$ and $j(x,t)=\rho(x,t)v(x,t)$ are the density and current of the Fokker-Planck Smoluchowski equation
\begin{equation}\label{eq:FP}
\frac{\partial \rho}{\partial t}=
-\frac{\partial j}{\partial x}=
-\frac{\partial}{\partial x}\left[
\beta D(x) f(x)\rho-D(x)\frac{\partial \rho}{\partial x}
\right]
\end{equation}
with $f(x)=-dF/dx$ the mean force.
Crucially, $v(x,t)$ must be estimated as a finite difference 
conditioned to be centered on $x$ at time $t$:~\cite{Seifert12}
\begin{align}\label{eq:Seifert2}
v(x,t)
\equiv\langle\dot{x}|x,t\rangle
=\lim_{\Delta t\to 0}\frac{
\langle [x(t+\Delta t)-x(t-\Delta t)]|x(t)\equiv x\rangle
}{2\Delta t}
\end{align}

In the numerical applications we are forced to adopt a finite $\Delta t$:
to ensure that the analysis is performed in the overdamped regime, as discussed in Supplemental Material, $\Delta t$ must be set larger than the correlation time $\gamma^{-1}$ (with $D=k_BT/m\gamma$) of the order parameter velocity $\dot x$ (see also the discussion in  Ref.~\cite{Peters_book} section 15.2).
As usual in Kramers-Moyal's scheme, $2D(x)\Delta t$ is estimated as the variance of 
$x(t+\Delta t)-x(t)$ conditioned at $x(t)\equiv x$.


The key finding of this work is that the more a coordinate $x$ is similar to the committor, the larger the magnitude of Seifert's $\Delta S_{tot}$ estimated for the relaxation from a barrier top, until reaching the true entropy production value (thus $-\beta\Delta F$) when $x$ is the committor $\varphi$ or any invertible function of the latter.
As we will show, this translates into a numerically efficient variational principle.
Compared to Ref.~\cite{Louwerse22},
where the committor was shown to preserve the whole production of the system's part only of the entropy, based on an analysis of long ergodic trajectories, in this work we fully account also for the heat dissipation part of the total thermodynamic entropy, and we introduce a straightforward numerical approach -- based on short relaxation trajectories -- to both coordinate optimization and free-energy calculation.

As a preliminary illustration, we consider a one-dimensional double-well model (Fig. 2). Hundreds of Langevin trajectories are initiated at the barrier top ($x=0$) and allowed to evolve until reaching local equilibrium in either well (see SI for details).
%
\begin{figure}
    \centering
    \includegraphics[width=0.5\textwidth]{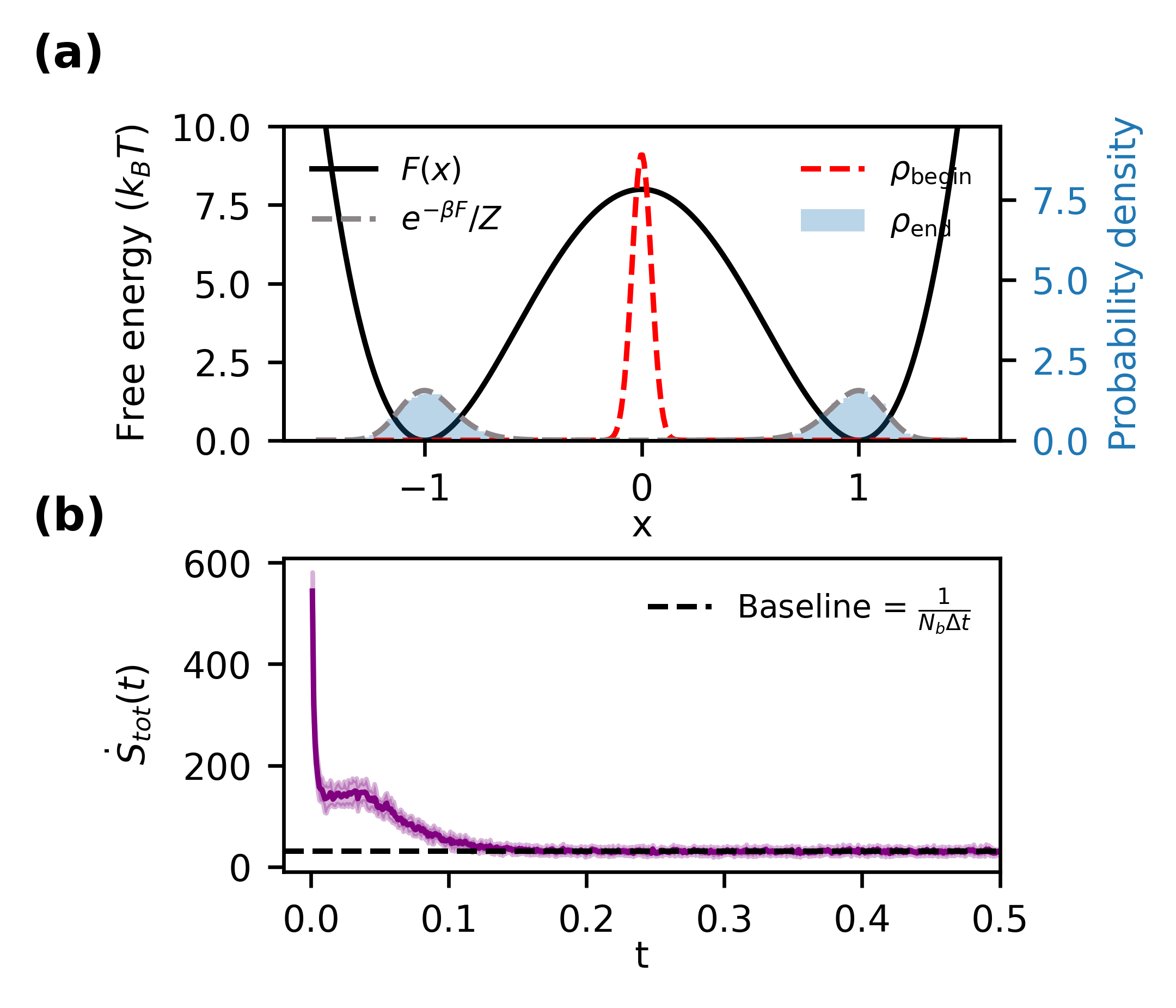}
    \caption{(a) One-dimensional free energy landscape $F(x) = 8(x^2 - 1)^2$ with the initial and final distributions (compared with the equilibrium density) of Langevin trajectories initiated at $x=0$. (b) EPR estimated from the trajectories. the darker line and the shaded area represent, respectively, the mean and the standard deviation computed over 20 independent blocks of 1000 trajectories.}
\end{figure}
%
\begin{figure*}[!ht] 
    \centering
    \includegraphics[width=0.9\textwidth]{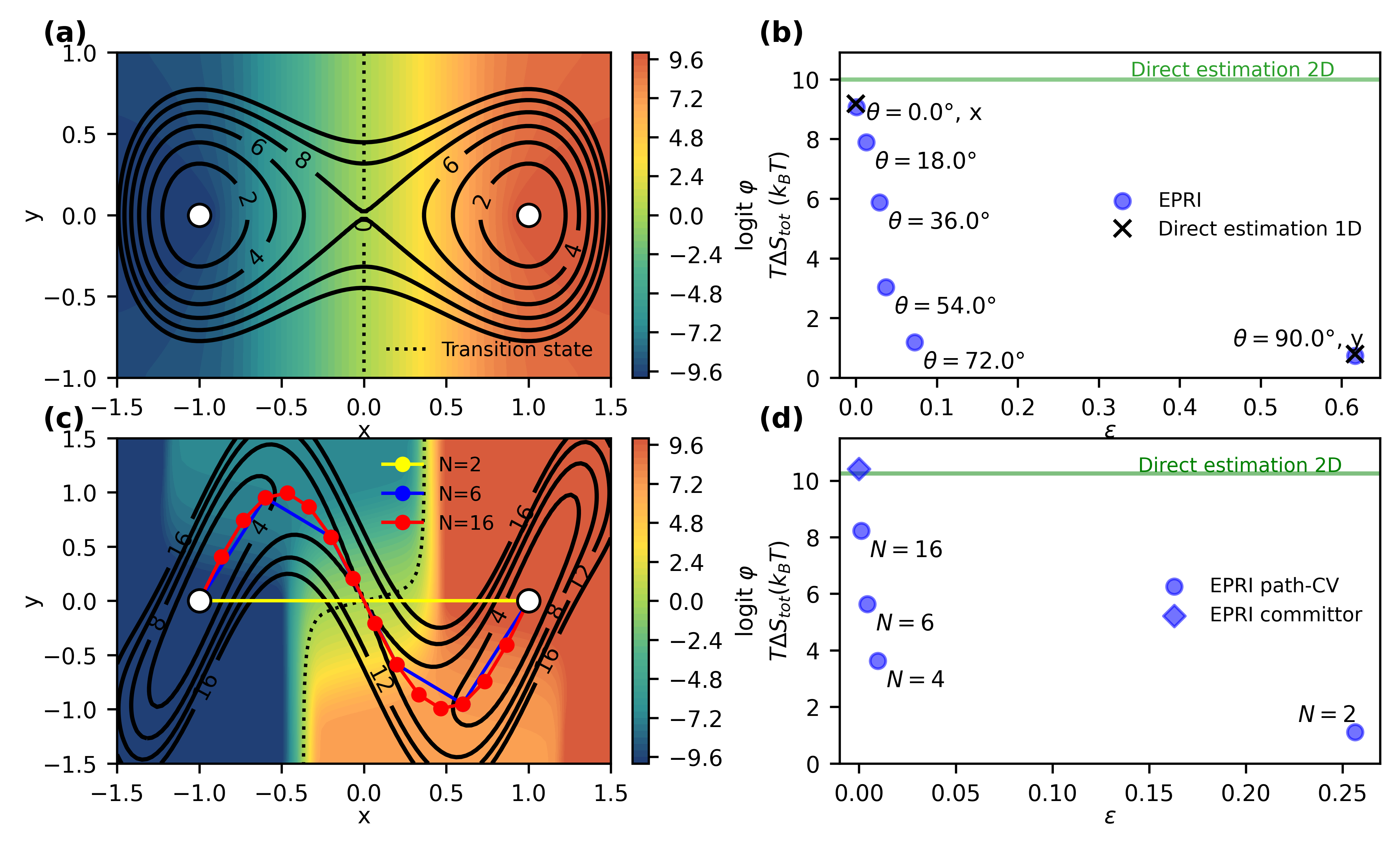}
    \caption{(a) 
    Double well free-energy landscape with committor $\varphi(x,y)$ aligned with the $x$ direction (see Supplemental Material for details). Free-energy isolines are in black, while the color-scale refers to the committor. (b) Total entropy production $\Delta S_{tot}$, estimated with the EPRI approach over $2 \cdot 10^4$ trajectories relaxing from the transition state, and distance $\epsilon$ from the committor for a set of linear CVs forming an angle $\theta$ with the $x$ direction. Reference values, indicated by black crosses and the green line, are obtained from direct estimation of $-\beta \Delta F=\log(P_{end}/P_{begin})$.
    (c) $N$-shaped double well with three path-CVs of increasing number of nodes $N$ (see Supplemental Material for details).
    (d) Total entropy production as in panel b, for path-CVs of different accuracy. The diamond corresponds to the coordinate logit$(\varphi)=\log[\varphi/(1-\varphi)]$. All error bars are $<0.25~k_BT$.
    }
\end{figure*}
Note that, since the velocity estimator eq.~\ref{eq:Seifert2} is based on a centered finite difference, the effective initial distribution is the one evaluated at $t=\Delta t$, i.e., a narrow Gaussian with standard deviation $\sqrt{2D \Delta t}$.  
By construction, the analysis can only be performed along $x$, all other degrees of freedom being hidden into the bath.

After discretizing space and time on grids, the EPRI in eq.~\ref{eq:Seifert1} is estimated from the relaxation trajectories.
We remark that, once local equilibrium is reached, $\dot{S}_{tot}$ does not vanish but it fluctuates around a baseline, as expected from theory (see Supplemental Material): the baseline can be systematically removed to estimate the EPRI value, $\Delta S_{tot}^x = 9.10 \pm 0.09 k_BT$. 
The latter matches precisely the direct estimation of entropy production as the exact expression 
$-\beta\Delta F = \log(P_{end}/P_{begin})=9.19\, k_BT$ 
computed via numerical integration over the free-energy landscape: note that between 100 and 1000 trajectories are sufficient for an accurate estimation (see Supplemental Material).

In this example, given the height of the barrier in the landscape $F(x)$, the entropy dissipated into the thermal bath upon relaxation $T\Delta S_{bath} = \int dx \rho_{begin} F(x) - \int dx \rho_{end} F(x) = 7.42\, k_B T$ is much larger than the system's entropy increase due to the spread of the $x$-distribution,
$T\Delta S=-\int dx\,\rho_{end}\log\rho_{end}+ \int dx \rho_{begin}\log\rho_{begin}
= 1.77\, k_BT$.

As a second benchmark, we consider a two-dimensional system in contact with a thermostat. 
We analyze trajectories relaxing from the barrier top of a double well landscape $F(x,y)$,
and we estimate entropy production by projecting the dynamics on a one-dimensional coordinate $q$,
leading to a clear demonstration of how EPRI values depend on the choice of the observation coordinate.
Note that, in any multidimensional system, analyzing the dynamics of a single variable $q$ with a Langevin model implies assuming that the orthogonal degrees of freedom equilibrate quickly compared to $q$.

In Fig. 3a, the optimal reaction coordinate is clearly $x$, i.e., the committor $\varphi(x)$ is a function of $x$ only, while $y$ does not carry information about the transition mechanism.
We recall that a generic coordinate $q$ is a function $\phi(x,y)$ and yields a one-dimensional free-energy profile
\begin{equation}
\beta F(q)=-\log\left[
\int dx \int dy\,\frac{e^{-\beta F(x,y)}}{Z}\delta\big(q-\phi(x,y)\big)
\right]
\end{equation}
As discussed above (see also Fig. 1), one expects $F(q)$ to typically overestimate the probability of the transition state ensemble, as only the committor can resolve it perfectly from other states, thus $\Delta S_{tot}^q \leq \Delta S_{tot}^\varphi$.

Fig. 3b shows the EPRI values for a family of coordinates $q = x\cos(\theta) + y\sin(\theta)$: when transforming from the $y$ to the $x$ direction (i.e., for $\theta$ decreasing from $90^\circ$ to zero),  $\Delta S_{tot}^q$ starkly increases, until reaching a value close to the true $\Delta S_{tot}$ of the bi-dimensional relaxation process, thus corroborating the variational nature of the EPRI.
An equivalent trend can be observed in Fig. 3c,d, where the transition path connecting the two wells has now been deformed to a highly-non linear N-shaped curve. 
Here, we consider a family of curvilinear coordinates, defined as path collective variables~\cite{branduardi07} with increasing number of reference points, that approximate more and more closely the committor $\varphi(x,y)$. 
Also in this case, clearly, the more similar to the committor is a coordinate, the larger is the fraction of the true entropy production that is captured by the EPRI. 

Finally, to test the new approach on a realistic complex system, we analyze a pair of C$_{60}$ fullerene nanoparticles forming a complex in water solution at 300~K and 1~atm.
As discussed in previous works~\cite{Palacio22},
this system presents a transition state for the  association/dissociation of the complex at a distance of about 1.2~nm between the centers of mass of the fullerenes.
We generated MD trajectories, according to the protocol detailed in Supplemental Material, that spontaneously relax from a single transition state configuration towards the locally-stable dimer state (we discard trajectories leading into the dissociated state). 
We note that multiple initial conditions belonging to the same ensemble could be used as a more representative set.

\begin{figure}[!ht]
    \centering
    \includegraphics[width=0.5\textwidth]{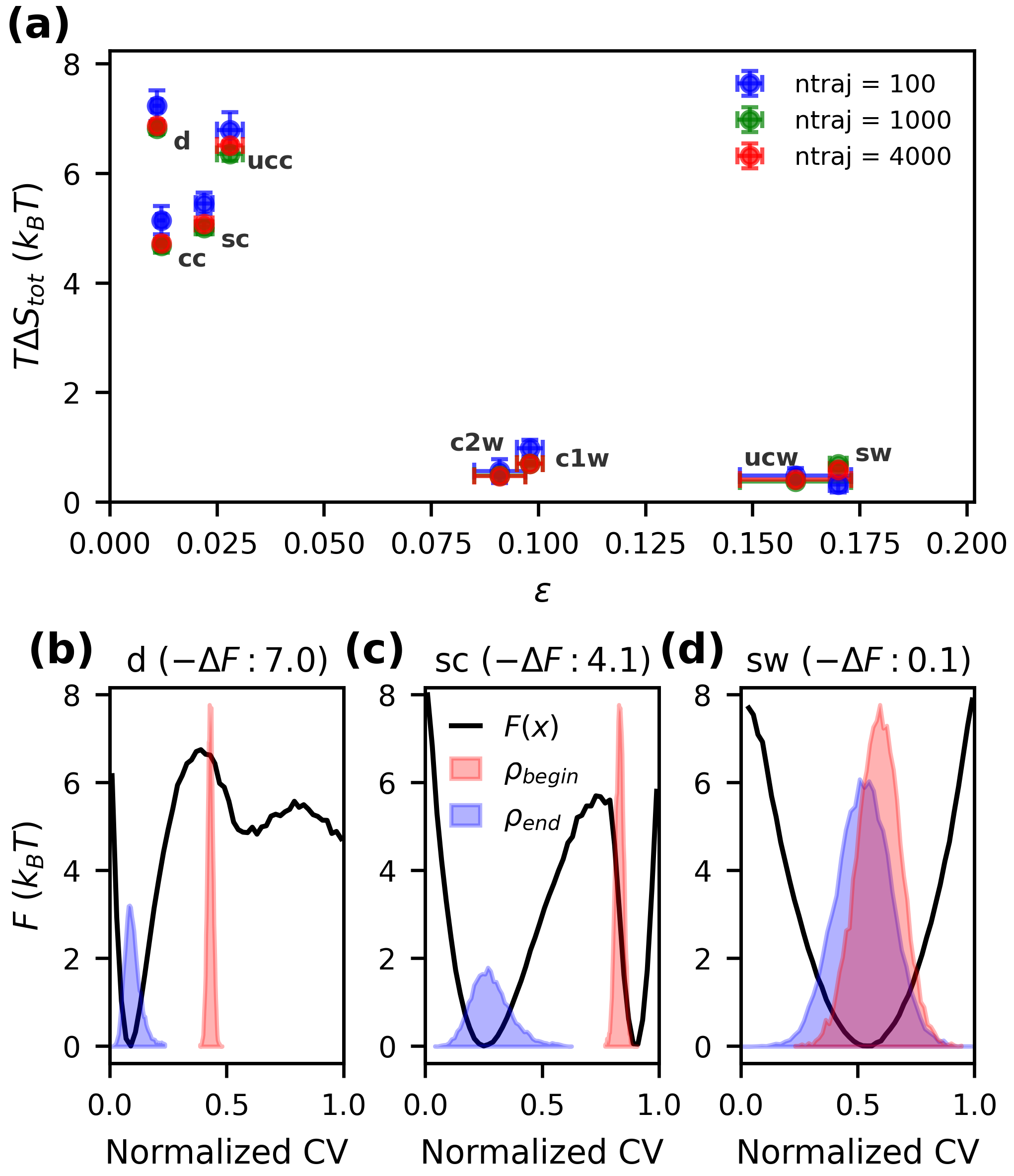}
    \caption{(a) Total entropy production $\Delta S_{tot}$ of fullerene dimer association trajectories relaxing from the transition state, as observed with different coordinates (see main text for definitions),
    and distance $\epsilon$ of the latter from the committor. Error bars are estimated using 5 blocks of trajectories. 
    Free-energy profiles from brute-force MD for three coordinates: (b) $d$, (c) $sc$ and (d) $sw$.
    Coordinates are shifted and scaled between 0 and 1.
    The initial and final probability densities of the relaxation process are shown in red and blue, respectively. The free-energy drop estimated directly as 
    $-\beta\Delta F=\log(P_{end}/P_{begin})$ from $F(q)$ integration (See Supplemental Material for details) is also indicated.} 
\end{figure}

As shown in Fig. 4, we consider a set of 8 putative transition coordinates able to describe, to a different extent, the mechanism of dimer formation.
In short, $d$ is the distance between the centers of mass of the $C_{60}$ molecules, $cc$ is the number of intermolecular contacts, 
$c1w$ and $c2w$ are the numbers of contacts between one molecule or both molecules and the solvent, $ucc$ and $ucw$ are the fullerene-fullerene and fullerene-solvent potential energies, and, finally, $sc$ and $sw$ are approximate two-body entropies of the solutes or of the solvent, respectively (see Supplemental Material for precise definitions). 

When comparing, for each coordinate, the similarity to the committor (estimated from a regression approach, as detailed in Supplemental Material) versus the EPRI magnitude, we find again a significant correlation.
The best coordinates can be clearly identified as those 
with the largest $\Delta S_{tot}^q$ values.
Moreover, the latter values approach the barrier height $-\beta\Delta F^q$ values estimated from the profile $\beta F(q)=-\log \rho(q)$ of brute-force MD,
an important confirmation that maximizing the EPRI leads towards optimal coordinates {\sl and} accurate free-energy estimations, circumventing the need for traditional free-energy calculation methods. 

It is important to remark that EPRI estimation requires a very limited computational effort: about 100 MD trajectories of 30 ps allow ranking the quality of the different coordinates and even estimating the free-energy barrier with moderate error bars compared to 40 times more data (Fig.~4).

Perhaps counter-intuitively, even though the complex association process involves a reorganization of the water molecules in the solvation shells of the carbon nanoparticles, the coordinates that best capture the transition mechanism are the "dry" ones, i.e., not explicitly including the water degrees of freedom.
We remark that a result consistent with the present one was found in Ref.~\cite{Mouaffac23}, when applying another criterion of reaction coordinate quality, namely based on the following variational principle: projecting a high-dimensional transition process on (any invertible function of) the committor, compared to any other coordinate, leads to a Langevin model with the slowest possible kinetics, faithfully reproducing the high-dimensional one~\cite{Zhang16}.

In fact, we argue that the two optimization criteria, i.e., maximizing the EPRI or minimizing the kinetic rate as a function of the coordinate $q$, are in fact equivalent: first of all, because both criteria lead to the committor, and second, more intuitively, since maximizing the entropy production leads to the lowest possible ratio $P_{TS}/P_B$ (as explained above), which is the primary factor slowing down the kinetic rate for an overdamped Kramers problem~\cite{Peters_book,Mouaffac23}.

Clearly, EPRI maximization is also equivalent to other reaction coordinate optimization methods in the literature, as long as they posit the committor as target, including Refs.~\cite{Peters06,Jung23,Kang24,Megias25}.
However, we underline the simplicity and reduced computational cost of our approach, able to provide a straightforward loss function for future machine learning algorithms.

Conceptually, the present approach focuses the attention on the fact that, for a closed system, rare events always originate from free-energy barriers that are purely entropic, when we acknowledge both the contributions of the system and of the thermal bath (see Eq.~\ref{eq:Stot_def}).
In this light, barrier-climbing is a rare event since it defies, at the microscopic scale, the second law, and entropy-production methods appear fully appropriate to address the problem, as illustrated in this article. 


Our work provides a numerically convenient technique (the code is freely available at \url{https://github.com/physix-repo/EPR/}) to free-energy estimation and order parameter optimization based on the second law of thermodynamics.
It seems to us that the theory discussed herein sheds light on the deep meaning and equivalence of available methods to obtain optimal order parameters, pointing to an overarching interpretation of the best reaction coordinate as the most irreversible one. 
This viewpoint, in addition, exposes the tight connection between Boltzmann's focus on the central problem of relaxation to equilibrium, and Onsager's intuition introducing the committor as order parameter~\cite{Onsager38}.

Future work is desirable to rigorously extend this approach also to the inertial and to the non-Markovian Langevin regimes, to broaden significantly the spectrum of potential applications.
Furthermore, including an analysis of the work of an external force, irrelevant in this paper but readily available in the theory of stochastic thermodynamics, could enable new reaction coordinate scoring tools in the field of enhanced sampling, where rare transitions are accelerated by artificial biasing forces.

\section{Acknowledgements}

We gratefully acknowledge Léon Huet and David D. Girardier for insightful discussions, and Hadrien Vroylandt for the computer code to estimate the committor function in the two-dimensional benchmarks.
Calculations were performed on the GENCI-IDRIS 
French national supercomputing facility, 
under Grant No. A0190811069.


\bibliographystyle{apsrev4-2}
%


\clearpage

\onecolumngrid
\appendix
{\LARGE\bf{Supplemental Material}}

\section{Direct estimation of the exact entropy production from the free-energy landscape}

First, we consider a one-dimensional system in contact with a heat bath, described with an overdamped Langevin model
\begin{equation}
    q(t+\Delta t) = q(t) -\beta D\frac{dF}{dq}\Delta t 
    + \sqrt{2 D \Delta t}\, R(t)
\end{equation}
with $F(q)$ the free-energy profile, and $R(t)\sim \mathcal{N}(0,1)$ a Gaussian random number
(note that, when $D$ is position dependent, an extra term $(dD/dq)\Delta t$ needs to be included).
We consider the relaxation between two local-equilibrium macrostates, $A$ and $B$, that correspond to normalized probability densities $\rho_A(q)$ and $\rho_B(q)$:
the following results apply also to the case where the initial density is, in reality, obtained from a set of shooting configurations in the transition state ensemble, 
since this situation can be treated as a constrained equilibrium state, for the sake of theoretical treatment.
Probabilities and free energies are given by
$$
\beta\Delta F=-\log\frac{P_B}{P_A}=-\log\frac{Z_B}{Z_A}
$$
$$
-k_BT\log Z_A=-k_BT\log \left(
\int_A dq\, e^{-\beta F(q)}
\right) = F_A
=\langle U\rangle_A-TS_A
$$
and likewise for $B$, where $U$ is the potential energy (the kinetic energy contribution to $\langle\Delta E\rangle$ cancels out at constant $T$).
In the main text, one- or two-dimensional Langevin models
are conceived as effective models of a high-dimensional system, typically simulated with molecular dynamics, so that we systematically replace the  potential energy notation with the free energy landscape notation:
$$
\langle U\rangle_A-TS_A \equiv \langle F\rangle_A-TS_A
=\int dq\,\rho_A(q)\,F(q) 
+ k_BT\int dq\,\rho_A(q)\,\log \rho_A(q)
$$
the last expression corresponds indeed to $-k_BT\log Z_A$, as it is easy to verify by replacing
$\log\rho_A(q)=\log\left( e^{-\beta F(q)}/Z_A \right)$.
The same treatment can be applied to $B$.
In this work, the direct estimation of the exact total entropy production $\Delta S_{tot}$ for the relaxation from $A$ to $B$, to obtain a reference value, is performed by numerical integration:
\begin{equation}\label{eq:SIdirect1}
    \Delta S_{tot}/k_B = -\beta\Delta F =
    \log
    \frac
{ \int dq\,\rho_B(q)\,\big[\,F(q) + k_BT\,\log \rho_B(q)\,\big] }
{ \int dq\,\rho_A(q)\,\big[\,F(q) + k_BT\,\log \rho_A(q)\,\big] }
\end{equation}
Clearly, this approach requires the knowledge of the free-energy landscape $F(q)$, that is not available in many real applications.

Next, we consider the case of a multi-dimensional system in contact with a thermal bath. For simplicity of notation, we take a two-dimensional system of coordinates $x,y$, so that the potential energy and the macrostate densities are now functions of two variables: 
\begin{align}
F_A=-k_BT\log Z_A & =-k_BT\log \left(
\int_A dx\,dy\, e^{-\beta U(x,y)}
\right)
=\langle U\rangle_A-TS_A\equiv \langle F\rangle_A-TS_A \\
& = \int dx\,dy\, \rho_A(x,y) [ F(x,y)
+k_BT \,\log \rho_A(x,y) ]
\end{align}
and likewise for state $B$. The direct estimate of the exact total entropy production becomes 
\begin{equation}\label{eq:SIdirect2}
    \Delta S_{tot}/k_B = -\beta\Delta F =
    \log
    \frac
{ \int dx\,dy\, \rho_B(x,y)\,\big[\,F(x,y) +k_BT \,\log \rho_B(x,y)\,\big] }
{ \int dx\,dy\, \rho_A(x,y)\,\big[\,F(x,y) +k_BT \,\log \rho_A(x,y)\,\big] }
\end{equation}
Eqs.~\ref{eq:SIdirect1},\ref{eq:SIdirect2} are employed to obtain reference values, to be compared with the EPRI approach of the next section.
We remark that, analytically, the same result is obtained from the definition of the partition functions:
$$
\Delta S_{tot}/k_B = \log 
\frac{ \int_B dx\,dy\, e^{-\beta F(x,y)} }
         { \int_A dx\,dy\, e^{-\beta F(x,y)} }
$$
however, for numerical calculations, this route can be more cumbersome (due to the need to identify the domains of the integrals) and it has not been followed in this work.

Finally, it is important to note that, when projecting a multi-dimensional system onto a single coordinate $q$, while at the same time considering the relaxation from a transition state ensemble to a local equilibrium state, the total entropy production evaluated in $q$-space generally underestimates the true value, as discussed in the main text,
since the projection artificially increases the probability of the transition state by mixing it with other states (unless the committor is used as projection coordinate):
\begin{equation}\label{eq:SIdirect3}
    \Delta S_{tot}^q/k_B = 
    \log
    \frac
{ \int dq\, \rho_B(q)\, \big[\, F(q) + k_BT\,\log \rho_B(q) \,\big] }
{ \int dq\, \rho_{TS}(q)\, \big[\, F(q) + k_BT\,\log \rho_{TS}(q)\, \big] }
 \leq \Delta S_{tot}/k_B
\end{equation}

\section{Velocity-based entropy production rate (EPR) estimator}

We consider the analysis of $N_{traj}$ trajectories of an order parameter $q$ relaxing towards local equilibrium. 

First, the order parameter position $q_j$ and time $t_k$ are discretized on grids, 
$q_j=j\Delta q$, $t_k=k\Delta t$,
and the position-dependent diffusion coefficient is estimated, based on Kramers-Moyal expansion, as the time-averaged variance of the forward displacement conditioned by the initial value:

$$
D(q_j) = \frac{\text{Var}(q(t+\Delta t) - q(t) \,|\, q_j\leq q(t)<q_{j+1})}{2\Delta t}
$$

A key quantity to estimate is the velocity field $v(q, t)$ defined by Eq.4 in the main text. To improve numerical performance, we use an adaptive spatial grid, instead of the previous regular one, to estimate the velocity field. At a given time $t_k$, sample points are equally distributed into $N_{bins}$ bins. To do so, we sort the $N_{traj}$ sample points according to their position values, and partition the points into $N_{bins}$ contiguous segments. Each segment defines an adaptive spatial bin $b$, centered at $q_b$, containing  $N_b = N_{traj}/N_{bins}$ sample points. In practice, the hyperparameters used are the number of trajectories $N_{traj}$ and the number of points per bin $N_b$.

We compute the velocity field in each bin $b$ as: 
\begin{equation}
   \bar{v}(q_b, t_k) = \frac{1}{N_b} 
   \sum^{N_b}_{
     \substack{\alpha = 1 \\ q_{\alpha}(t_k) \in b }
   }\frac{q_{\alpha}(t_k+\Delta t) - q_{\alpha}(t_k-\Delta t)}{2\Delta t}
   \label{eq:v_bin}
\end{equation}

Finally, the entropy production rate at time $t_k$ is computed as the following spatial average
(corresponding to the integral over $\rho(x,t)$
in Eq.~2 in the main text, given the uniform sample density across all bins): 
\begin{equation}
    \dot{S}_{tot}(t_k) = \frac{1}{N_{bins}} \sum_{b=1}^{N_{bins}}  \frac{\bar{v}^2(q_b, t_k)}{D(q_b)}
    \label{eq:sdot}
\end{equation}

where $D(q_b)$ is assigned as $D(q_j)$ with $q_j \leq q_b(t_k) < q_{j+1}$.
Time-integration until relaxation is completed provides the total entropy production.

\section{EPR estimator bias}    

In this section, we will show that operating at finite $\Delta t$ with finite data induces a deterministic bias in our EPR estimator (\ref{eq:sdot}). Fortunately, this bias takes a very simple form which can be subtracted to construct an unbiased estimator. We follow Seifert's derivation \cite{Seifert12}, with the exception that we do not take the $\Delta t \longrightarrow 0$ limit. We begin by defining the finite-$\Delta t$ estimator $\hat{v}(q,t)$ for the phase-space velocity $v(q,t)$ computed on a single trajectory such that $q(t) = q$:

\begin{gather}
    \hat{v}(q,t) = \dfrac{q(t+\Delta t) - q(t-\Delta t)}{2 \Delta t} \nonumber \\
    \hat{v}(q,t) = \dfrac{\Delta q^+ + \Delta q^-}{2 \Delta t}
    \label{eq:vel_est}
\end{gather}

where $\Delta q^+ = q(t+\Delta t) - q(t)$ and $\Delta q^- = q(t) - q(t-\Delta t)$ denote the forward and backward position increments, respectively. We recall that we work under the hypothesis that $q(t)$ obeys an overdamped Langevin equation with a drift function $\phi(q)$ and a diffusion profile $D(q)$:

\begin{equation}
    dq(t) = \phi(q(t)) dt + \sqrt{2 D(q(t))} dW(t)
\end{equation}

where $W(t)$ is a standard Wiener process. We now expand the position increments for small $\Delta t$ using this dynamical evolution equation. The forward increment straightforwardly yields:

\begin{equation*}
    \Delta q^+ = \phi(q) \Delta t + \sqrt{2 D(q)} I_{W^+} + O(\Delta t^{3/2})
\end{equation*}

where the It$\bar{\text{o}}$ integral $I_{W^+} = \int_t^{t+\Delta t} dW(s)$ is a zero-mean Gaussian random variable with variance $\Delta t$. Following Seifert's conditioning framework \cite{Seifert12}, the backward increment with endpoint conditioning can be expressed as:

\begin{equation*}
    \Delta q^- = \left[ \phi(q) - 2 D(q) \, \partial_q \ln \rho(q,t) \right] \Delta t + \sqrt{2 D(q)} I_{W^-} + O(\Delta t^{3/2})
\end{equation*}

where $\rho(q,t)$ is the time-dependent probability density. The backward It$\bar{\text{o}}$ integral $I_{W^-} = \int^t_{t-\Delta t} dW(s)$ is also a zero-mean Gaussian random variable with variance $\Delta t$ which is statistically independent from $I_{W^+}$ since the corresponding time intervals do not overlap. Substituting these expansions back into the velocity estimator (\ref{eq:vel_est}) yields:

\begin{equation*}
    \hat{v}(q,t) = \phi(q) - D(q) \, \partial_q \ln \rho(q,t) + \sqrt{2 D(q)} \dfrac{I_{W^+} + I_{W^-}}{2 \Delta t} + O(\Delta t^{1/2})
\end{equation*}

From this expression, we can readily compute the expectation value and variance of this single-trajectory estimator. In the remainder of this section, all expressions are evaluated at leading order in $\Delta t$. Taking the ensemble average, we find that the stochastic integrals vanish, giving:

\begin{equation}
    \mathbb{E}[\hat{v}(q,t)] = \phi(q) - D(q) \, \partial_q \ln \rho(q,t) = v(q,t)
    \label{eq:E(v)}
\end{equation}

Since $\hat{v}(q,t)$ is the sum of two independent Gaussian random variables, its variance is simply the sum of their variances:

\begin{equation}
    \text{Var}[\hat{v}(q,t)] = \dfrac{2 D(q)}{4 \Delta t ^2} (\Delta t + \Delta t) = \dfrac{D(q)}{\Delta t}
    \label{eq:Var(v)}
\end{equation}

Hence, while $\hat{v}(q,t)$ serves as an unbiased estimator of the local velocity $v(q,t)$, its variance goes to infinity as $\Delta t$ goes to 0. This property will determine the bias of our EPR estimator. 

We now turn back to our discrete description of the spatial grid and consider a bin $b$ centered on $q_b$. We study the statistical properties of $\bar{v}(q_b,t)$, i.e., the bin average of $\hat{v}$, as defined in eq.~(\ref{eq:v_bin}). 
This is an unbiased estimator of $v(q_b,t)$, with variance $\text{Var}[\bar{v}(q_b,t)] = \frac{D(q_b)}{N_b \Delta t}$ (neglecting the variations of $D$ within the bin).

Defining the contribution of bin $b$ to the EPR: $\bar{\dot{s}}(q_b,t) = \bar{v}^2(q_b,t) / D(q_b)$ and using the fundamental statistical identity $\mathbb{E}[X^2]=\mathbb{E}[X]^2+\text{Var}(X)$ as well as (\ref{eq:E(v)}) and (\ref{eq:Var(v)}), we obtain for $\mathbb{E}[\bar{\dot{s}}(q_b,t)]$:

\begin{equation*}
    \mathbb{E}[\bar{\dot{s}}(q_b,t)] = \dfrac{v^2(q_b,t)}{D(q_b)} + \dfrac{1}{N_b \Delta t}
\end{equation*}

\begin{figure}[!htbp]
    \centering
    \includegraphics[width=0.8\textwidth]{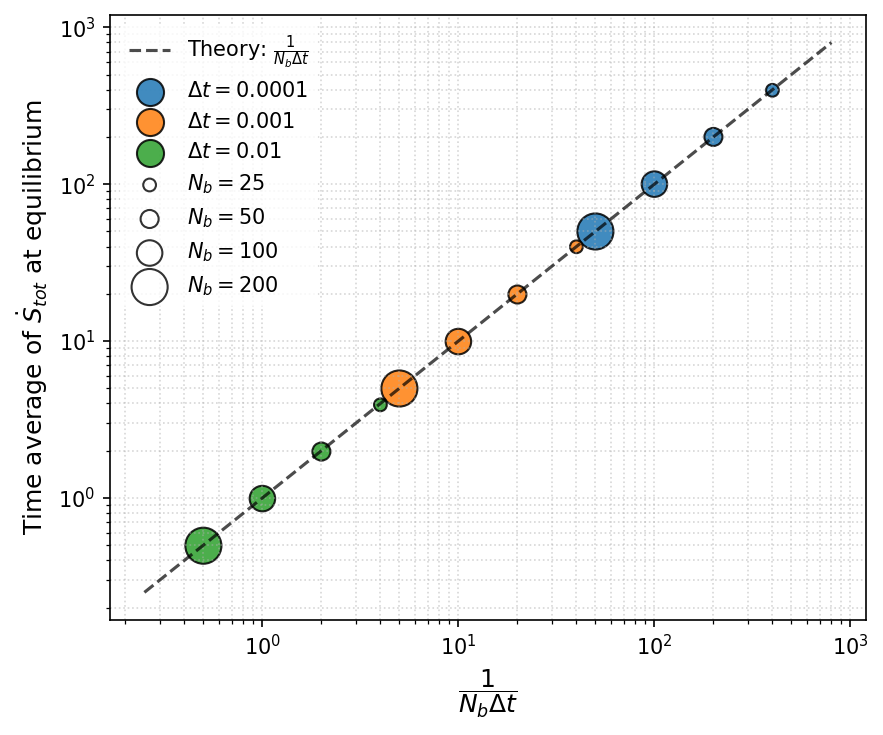}
    \caption{Measured entropy production rate of equilibrium trajectories, varying the number of points per bin $N_b$ and the timestep $\Delta t$. Data points represent the time average over 1500 steps of 10000 independent trajectories sampling the equilibrium ensemble of an Ornstein-Uhlenbeck process where $F(q) = - q$ and $D=1$. Because the true EPR at equilibrium is identically zero, the non-zero measured values stem entirely from finite-data bias, collapsing onto the predicted scaling law $\dot{S}_{tot} = 1/(N_b \Delta t)$.}
    \label{fig:epr_bias}
\end{figure}

Finally, averaging $\bar{\dot{s}}(q_b,t)$ over the uniformly distributed positions $q_b$ yields our EPR estimator $\dot{S}_{tot}(t)$ defined in (\ref{eq:sdot}). Its expectation value follows directly from that of $\bar{\dot{s}}(q_b,t)$:

\begin{equation}
    \mathbb{E}[\dot{S}_{tot}(t)] = \dot{S}_{tot}^{true}(t) + \dfrac{1}{N_{b} \Delta t}
    \label{eq:epr_bias}
\end{equation}

The EPR estimator has a bias which we proved to be equal to $1/(N_b \Delta t)$ at leading order in $\Delta t$, with $N_b$ the number of observations per bin and $\Delta t$ the time resolution of the trajectory. Numerical experiments on equilibrium trajectories (Figure \ref{fig:epr_bias}) confirm this result for different values of $N_b$ and $\Delta t$. It is worth noting that this bias vanishes in the limit of infinite data points. We recover the same form as Ref. \cite{Ronceray20} for the bias, up to a factor of 2; however, our own expression for the EPR estimator bias appears in full agreement with the aforementioned numerical benchmark. Note that the variance of the estimator is less straightforward to obtain, as outlined in Ref.~\cite{Ronceray20}, but it is not necessary in our case, as we can compute it empirically and we can minimize it by increasing the number of independent trajectories.

\begin{figure}[hb!]
    \centering
    \includegraphics[width=0.92\textwidth]{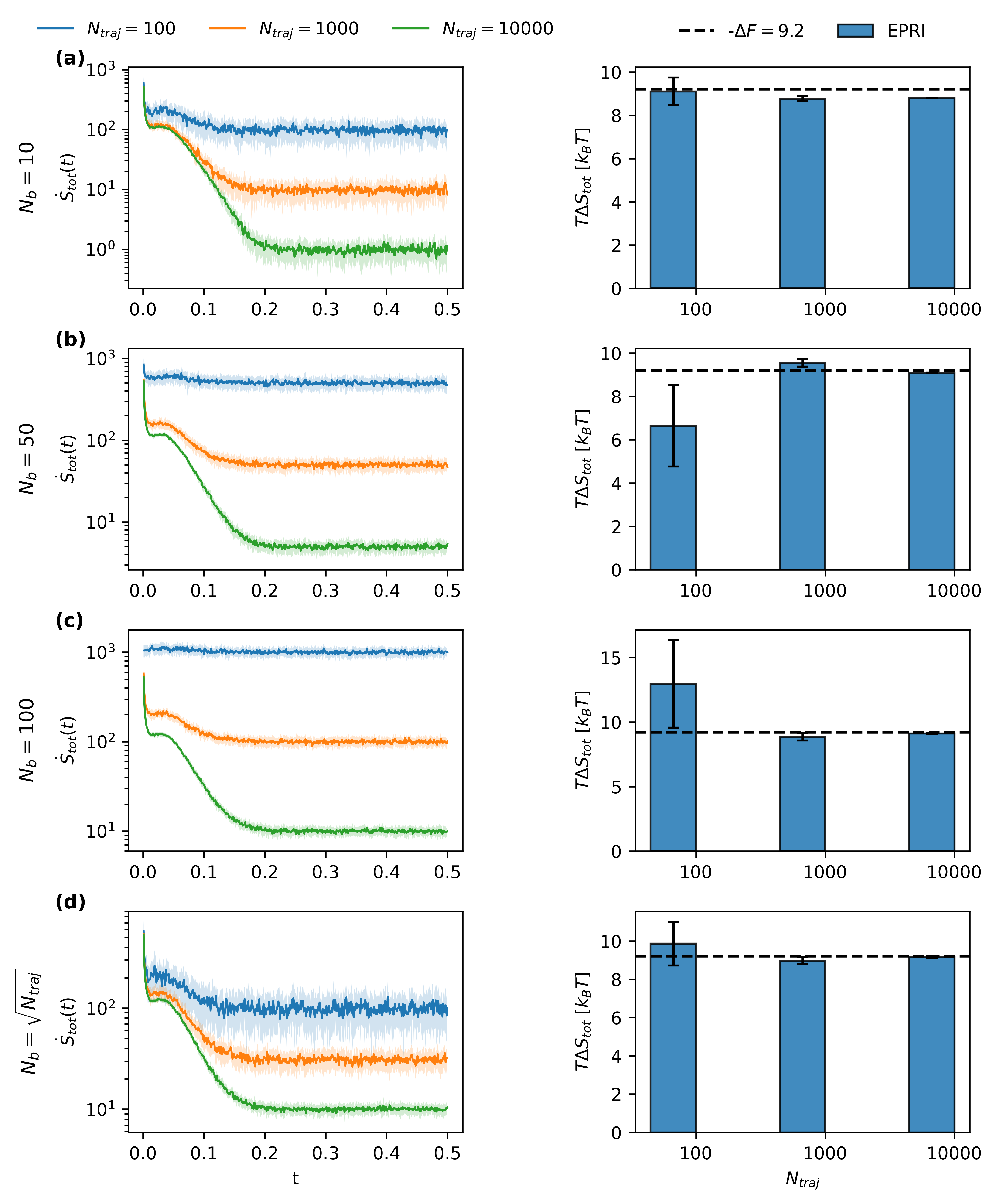}
    \caption{Influence of the parameter $N_b$ on the estimation of the total entropy production from shooting trajectories of the 1D double well of  Fig. 2 in the main text, with timestep $\Delta t = 0.001$. The exact $-\beta\Delta F=\Delta S_{tot}$ has been computed from direct integration as Eq.~\ref{eq:SIdirect1}. Based on these results,  $N_b = \sqrt{N_{ntraj}}$ has been taken as a good compromise for all the applications. Error bars have been computed over 20 independent blocks.}
    \label{fig:epr_ntraj}
\end{figure}

\clearpage
\section{Simulation details}

\subsection{Benchmarks}

We generated 20 independent sets of $10^3$ overdamped Langevin trajectories on a two-dimensional free energy landscape (of two different shapes, see below): 

\begin{align}
    x(t + \Delta t) - x(t) = & -\beta D\partial_x F(x, y)\Delta t + \sqrt{2D \Delta t} \,R_x(t) \\
    y(t + \Delta t) - y(t) = & -\beta D\partial_y F(x, y)\Delta t + \sqrt{2D \Delta t} \,R_y(t) \\
\end{align}

where $R_x$ and $R_y$ are independent Gaussian random numbers with zero average and unit variance. We integrated the former equations with $\beta=D=1$,
a timestep $\Delta t = 0.001$ for a length $t_{tot}=1$, from the initial condition $(x(0), y(0)) = (0, 0)$. Based on the analysis in Fig. 2, we set $N_b = \sqrt{N_{traj}}$. 

The double wells of main text Figures 3a and 3c have the following forms, respectively: 
$$F(x, y) = A(x^2 - 1)^2 + \frac{1}{2}K y^2$$
and
$$F(x, y) = A(x^2 - 1)^2 + \frac{1}{2}K \big[
y - \sin(\pi(x+1))
\big]^2$$
with $A = 8\,k_B T$ and $K = 40\,k_B T$. 
Collective variables $q(x,y)$ for the second, $N$-shaped landscape, have been defined as the "$s$" path-CV in Ref.\cite{branduardi07}, with a variable number $N$ of nodes uniformly distributed along the minimal free energy path connecting the two minima:
$$
q(x, y) = \frac{\sum_{k=1}^{N} k\, e^{-\lambda\big[(x-x_k)^2 + (y-y_k)^2\big]}}
{\sum_{j=1}^{N} e^{-\lambda\big[(x-x_j)^2 + (y-y_j)^2)\big]}}
$$
as usual in the literature, we set 
$\lambda = 2.3 / \frac{1}{N-1} \sum_{k=1}^{N-1} \big[(x_{k+1} - x_k)^2 + (y_{k+1} - y_k)^2\big]$
in order to obtain a smooth interpolation.

\begin{figure}
    \centering
    \includegraphics[width=0.6\textwidth]{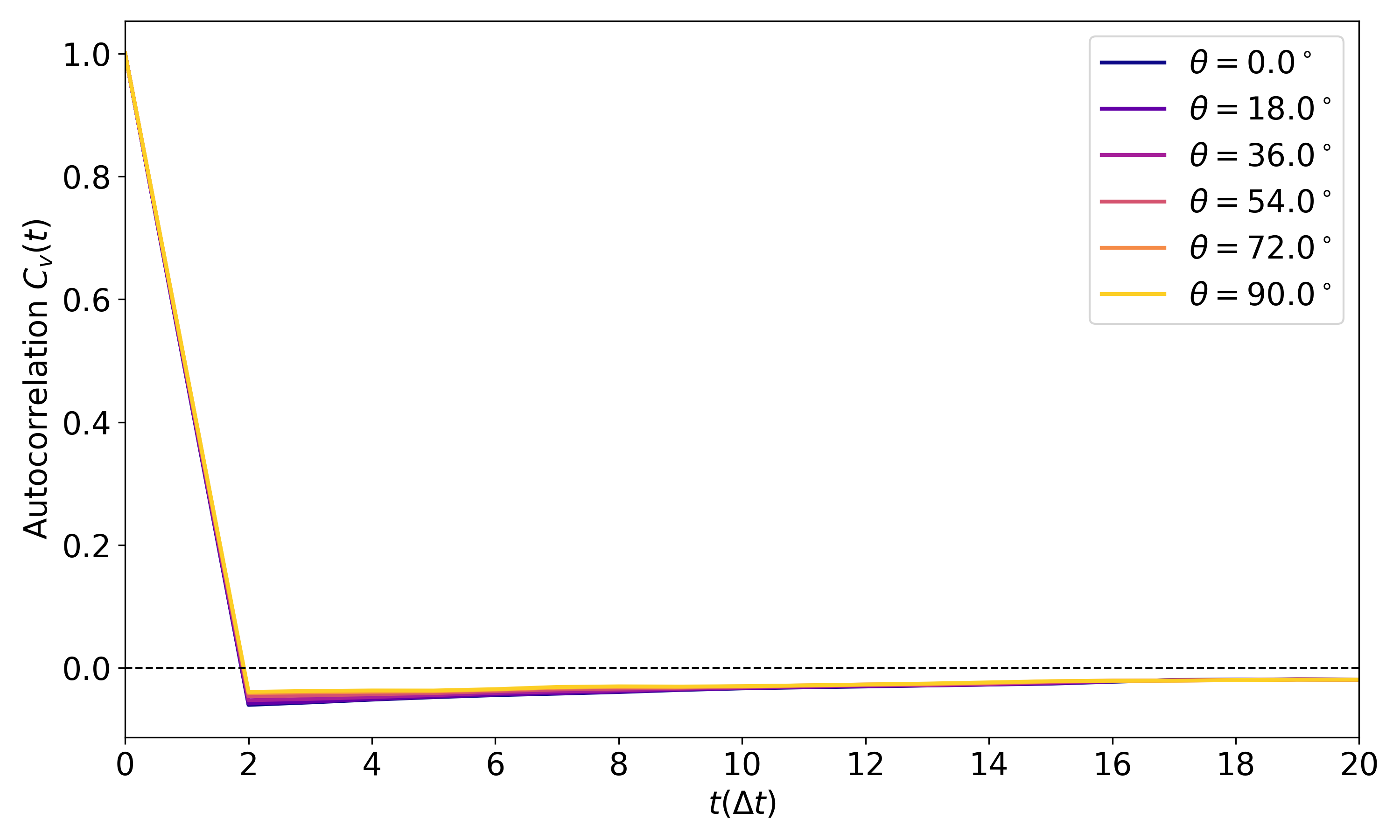}
    \caption{Normalized velocity autocorrelation function $\langle \dot{q}(t_0)\dot{q}(t_0+t)\rangle / \langle \dot{q}^2\rangle$
    for each CV used with the bidimensional double well oriented along $x$.}
    \label{fig:vacf_doublewell}
\end{figure}

\begin{figure}
    \centering
    \includegraphics[width=1.0\textwidth]{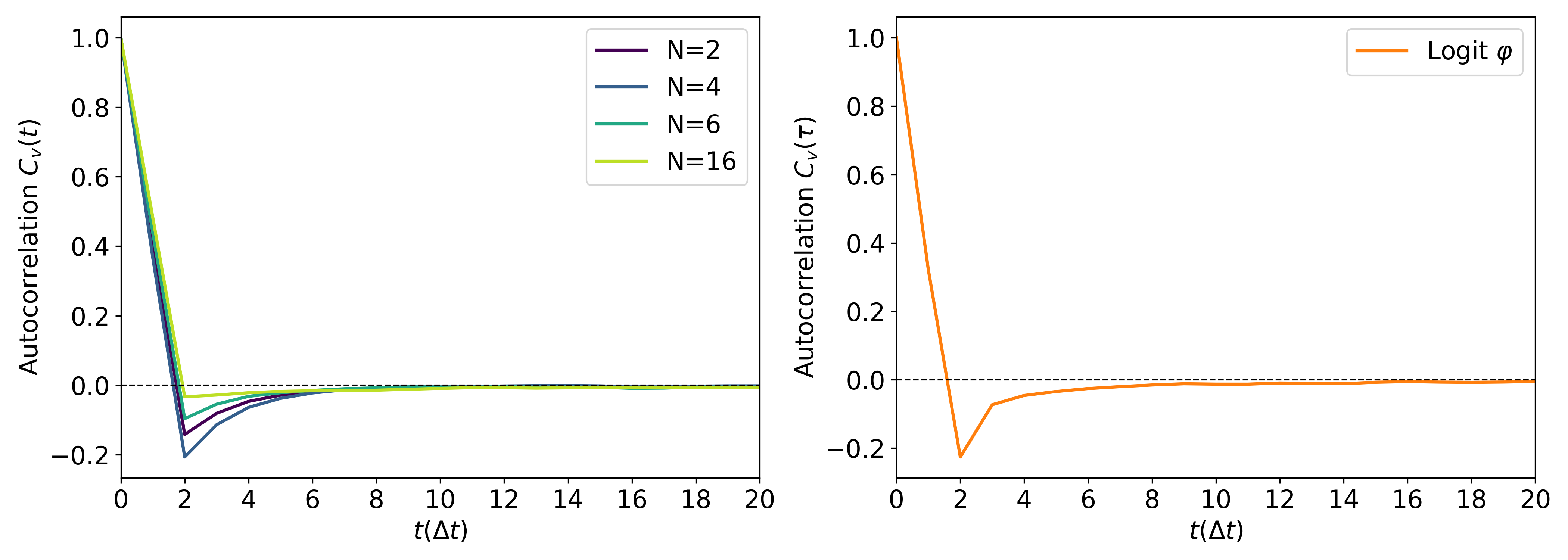}
    \caption{Normalized velocity autocorrelation function $\langle \dot{q}(t_0)\dot{q}(t_0+t)\rangle / \langle \dot{q}^2\rangle$
    for each CV used with the $N$-shape bidimensional double well.}
    \label{fig:vacf_nshape}
\end{figure}

For EPR calculations, we adopted a time resolution equal to $\Delta t$ (the integration time step here above) in the case of the first landscape, and equal to $5\Delta t$ for the second one, based on the velocity autocorrelation functions in Fig. ~\ref{fig:vacf_doublewell} and ~\ref{fig:vacf_nshape}, to ensure that the projected dynamics is approximately in the overdamped regime.

\subsection{Solvated fullerene dimer}

The association and dissociation of C$_{60}$ fullerene dimers (OPLS-AA force-field \cite{jorgensen1996opls}) in water solution (SPC force-field \cite{berendsen81}) have been simulated by MD using GROMACS v2019.4 \cite{Berendsen95, Abraham15} patched with PLUMED 2.5.3 \cite{plumed2}. 
Two C$_{60}$ molecules, whose initial coordinates and topology are taken from Ref. \cite{monticelli2012atomistic}, were solvated by 2398 water molecules in a simulation box of $3.607^3$~nm$^3$ with periodic boundary conditions. 

Geometry minimization exploited the steepest descent algorithm, stopped when the maximum force was $\leq 50$ kJ/mol$\cdot$ nm. 
We used the leapfrog algorithm to propagate the equations of motion and the non-bonding interactions were calculated using a PME scheme with a $1.2$ nm cutoff for the part in real space. 
A $100$ ps equilibration in the $NVT$ ensemble was performed with a  stochastic velocity rescaling scheme \cite{bussi2007canonical} with a time constant of 1~ps, followed by a 100~ps equilibration in the $NPT$ ensemble using the Parrinello-Rahman barostat \cite{parrinello1981polymorphic} with time constant of 4~ps and a time step of 1 fs.

MD production trajectories were generated without restraints, with a time step of 1~fs in the $NVT$ ensemble at 298~K with a thermostat time constant of 1~ps. 

First, a long ergodic trajectory of 200 ns was generated, with CVs saved every 1 ps,  to observe many recrossings between the associated and dissociated state, allowing to estimate the free-energy landscape $\beta F(q)=-\log\rho(q)$ for any CV $q$. 

Second, we shot $5\cdot 10^4$ trajectories of 30 ps each from a transition state configuration, with CVs saved every 0.1 ps, and selected, for further EPRI calculations, the 21~987 trajectories relaxing into the associated state.

\subsection{Collective variable (CV) definitions}

The following set of CVs was employed to project the fullerene dimer trajectories on a one-dimensional space:

\begin{enumerate}
    \item $d$: the distance between the centers of mass of each fullerene molecule.
    
    \item $cc$: the number of carbon-carbon contacts, estimated summing  continuous coordination functions between any atom of the first fullerene (set $S_1$) and any atom of the second fullerene (set $S_2$)  
    \begin{equation}\label{eq:ful-cvs-c}
    cc = \sum_{i \in S_1} \sum_{j \in S_2} C_{ij}\ ,\ \ \ 
     C_{ij} = \frac{1-\Big(\frac{r_{ij}}{r_0}\Big)^n}{1-\Big(\frac{r_{ij}}{r_0}\Big)^m}
    \end{equation}
    where $r_{ij}$ is the distance between atoms $i$ and $j$, with parameters $r_0=0.35$~nm, $n=6$ and $m=10$. 
    
    \item $c2w$: the number of carbon-water contacts, defined similarly to $cc$ in Eq.~\ref{eq:ful-cvs-c}, in this case with $r_0=0.6$~nm, set $S_1$ including all carbon atoms of the two fullerene molecules, and set $S_2$ including all oxygen atoms of the water molecules.

    \item $c1w$: the water-carbon contacts for a single fullerene molecule, defined as $c2w$ except for the inclusion of a single fullerene in set $S_1$. 
     
    \item $sc$: the approximate carbon pair entropy, estimated using the expression of Ref.~\cite{piaggi2017enhancing}
    \begin{equation}
    sc=-2\pi\rho\, k_B \int_0^{r_{\mathrm{max}}} \left [ g(r) \ln g(r) - g(r) + 1 \right ] r^2 \mathrm{d}r~,
    \end{equation}
    where $\rho$ is the average atomic density, $r_{\mathrm{max}}$ is a cutoff set to 0.65~nm, and $g(r)$ is the pair distribution function of carbon atoms, estimated via Gaussian kernels as
    \begin{equation}
    g(r) = \frac{1}{4 \pi N \rho\, r^2} \sum_{i\neq j} \frac{1}{\sqrt{2 \pi \sigma^2}} e^{-(r-r_{ij})^2/(2\sigma^2)}~,
    \end{equation}
    where $N$ is the number of carbon atoms and $\sigma = 0.01$~nm. The carbon and water pair entropies were calculated using PLUMED \cite{plumed2} (keyword: PAIRENTROPY). 
    
    \item $sw$: the approximate water pair entropy, estimated with the same equations and parameters as $sc$ applied to oxygen atoms.
    
    \item $ucc$: the Van der Waals carbon-carbon potential energy, computed over all inter-fullerenes carbon pairs. Note that this definition is different from the one in Ref.\cite{Mouaffac23}, where interactions within the same molecule are also included.
    
    \item $ucw$: the Van der Waals carbon-water potential energy, computed between all the carbon-solvent atom pairs.
\end{enumerate}

\begin{figure}
    \centering
    \includegraphics[width=1.0\textwidth]{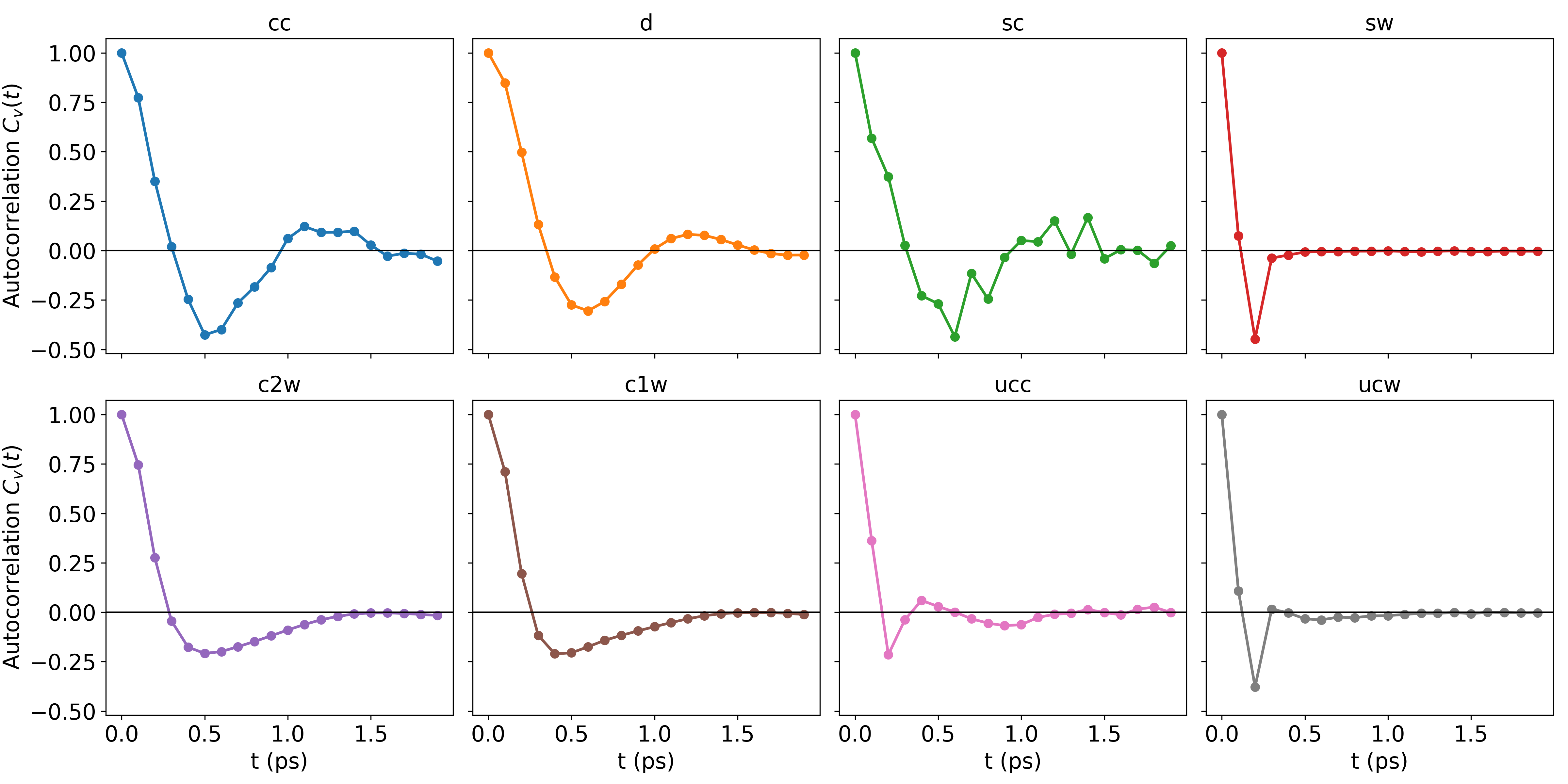}
    \caption{Normalized velocity autocorrelation function $\langle \dot{q}(t_0)\dot{q}(t_0+t)\rangle / \langle \dot{q}^2\rangle$
    for each CV used with the fullerene dimer system.
    The time resolution of the EPR analysis is chosen as $\Delta t = 1$~ps, so that the overdamped regime is a good approximation of the $q$-projected MD trajectories.}
    \label{fig:vacf}
\end{figure}

CV trajectories are analyzed with a time resolution $\Delta t=1$~ps for EPR calculations, based on the 
results of Figure~\ref{fig:vacf}.

\begin{figure}
    \centering
    \includegraphics[width=1.0\textwidth]{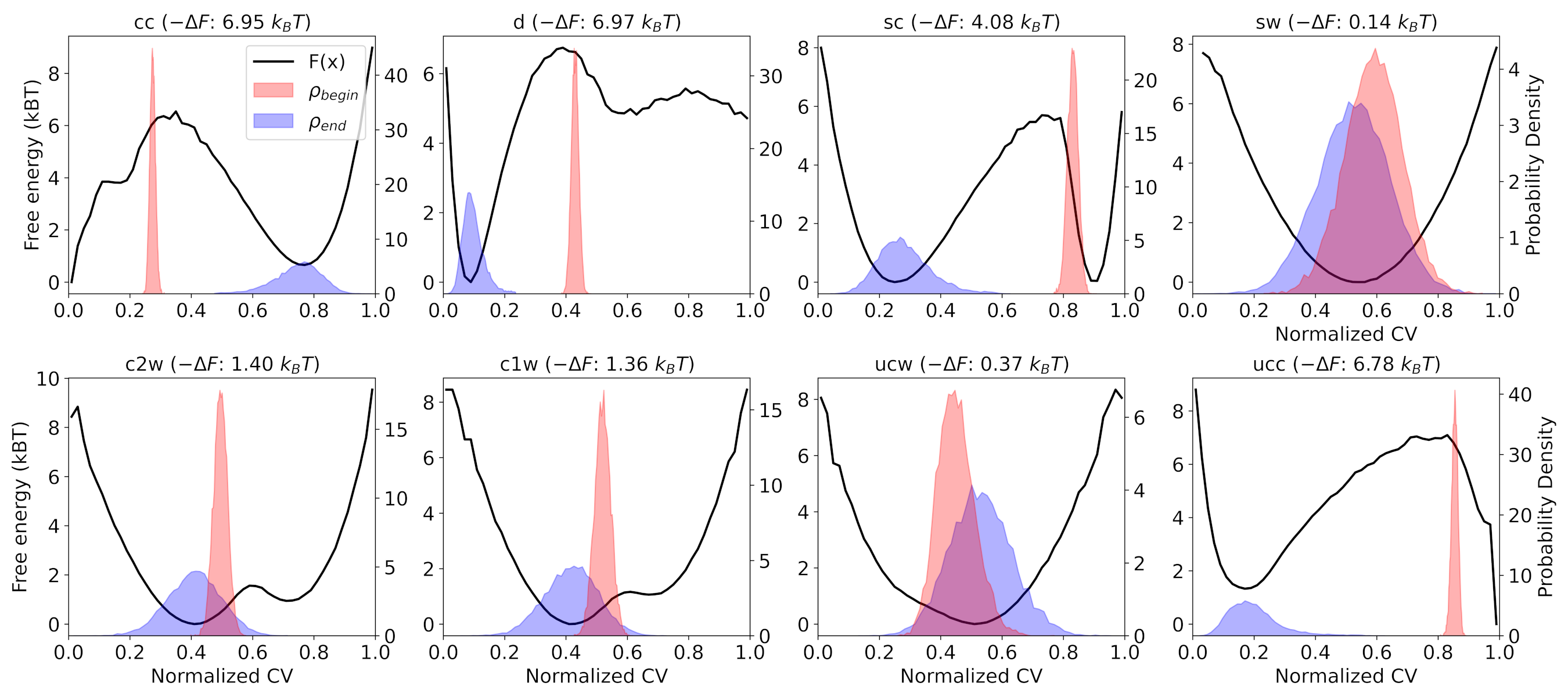}
    \caption{Brute-force free-energy profiles of the CVs employed to analyze the fullerene dimer in water. 
    CVs are shifted and scaled between 0 and 1.
    The initial and final probability densities of the relaxation process are shown in red and blue, respectively. The free-energy drop estimated from direct integration (Eq.~\ref{eq:SIdirect3}) is also indicated.}
    \label{fig:Fprofiles}
\end{figure}

Figure~\ref{fig:Fprofiles} shows the profiles $\beta F(q)=-\log\rho(q)$ obtained from  ergodic MD trajectories, together with the probability densities of the relaxation trajectories at $t=\Delta t$ and at the final time (local equilibrium in the associated state).
We remark that lower-quality CVs display a smaller barrier, with an initial density (in red) that can be displaced from the barrier top, despite being at the transition state with $\varphi=1/2$ (see, e.g., $sc$).

\begin{figure}
    \centering
    \includegraphics[width=0.8\textwidth]{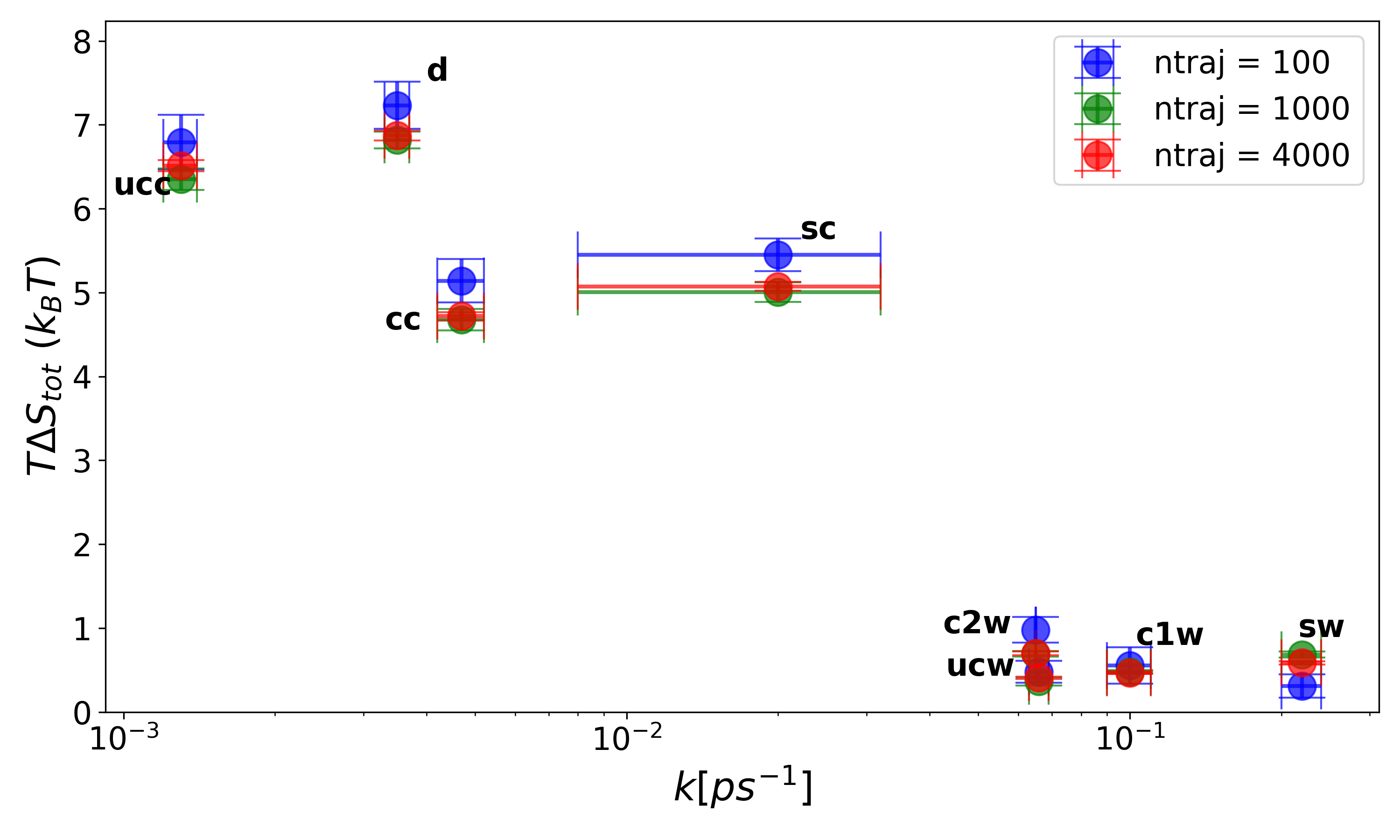}
    \caption{Total entropy production $T\Delta S_{tot}$ as a function of the estimated kinetic rate $k$ (inverse of the mean first passage time for dissociation) in the solvated fullerene dimer system. Kinetic rates are estimated from  overdamped Langevin models obtained by a maximum likelihood scheme, as detailed in Ref.~\cite{Mouaffac23}. Lower rates and higher entropy production indicate better CVs.}
    \label{fig:epr_k}
\end{figure}

\begin{figure}
    \centering
    \includegraphics[width=0.8\textwidth]{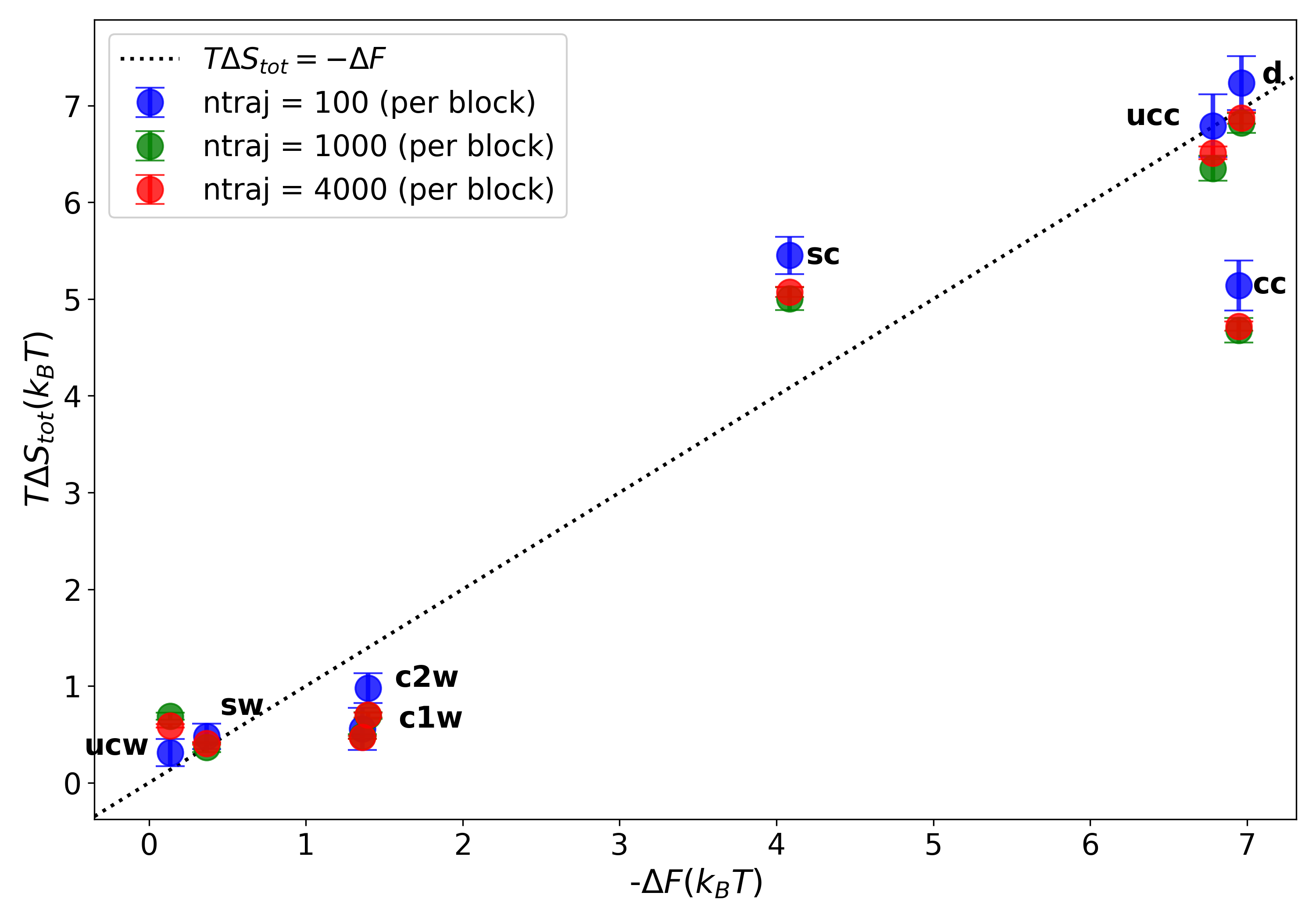}
    \caption{Total entropy production $T\Delta S_{tot}$ as a function of the direct integration of $-\Delta F$ computed with brute force free-energy profiles from Figure~\ref{fig:Fprofiles}.}
    \label{fig:epr_deltaF}
\end{figure}

\clearpage
\section{Estimation of the committor and distance definition}

For the one- and two-dimensional benchmarks, the committor function $\varphi$ with respect to two states $A$ and $B$ can be estimated with high precision by solving the following backward Kolmogorov equation:~\cite{Metzner06,Roux22}

$$
\nabla \cdot \left[ D\, e^{-\beta F(x, y)}\, \nabla \varphi(x, y) \right] = 0
$$ 

with boundary conditions 
$
\varphi|_{\partial A}=0,\  \varphi|_{\partial B}=1
$.
To this aim, we employed a code developed by Hadrien Vroylandt (Caen University) in the Folie package, available at \url{https://github.com/langevinmodel/folie/}.

In the case of molecular dynamics simulations of a complex system, the committor can be stimated for each selected $3N$-dimensional atomic configuration $\mathbf{r}$ by shooting a large-enough number of trajectories from it, with random, Boltzmann-distributed initial velocities, and counting how many of them hit state $B$ before state $A$. 
In the case of the solvated fullerene dimer, 100 trajectories have been shot from each configuration of a set of 300, uniformly distributed along the sidtance $d$ between the two fullerene centers of mass. The associated state is defined as $d<1.1$~nm, the dissociated state as $d>1.3$~nm. 

Given a collective variable $q$, we define the distance $\epsilon$ between $q$ and $\varphi$,
following Ref.~\cite{Mouaffac23},
as the residual sum of squares (normalized with respect to the number of configurations)
of a fit of the committor values $\varphi(q(\mathbf{r}))$ with the following sigmoid function:\cite{Peters06,Jung23}
$$
\Phi(q; a, b) = \frac{1}{1 + e^{-\frac{{q-a}}{b}}} = \frac{1}{2}\left[ 1 + \tanh \left( \frac{q-a}{2b} \right) \right]
$$



\end{document}